\documentclass[prx,superscriptaddress,twocolumn,longbibliography]{revtex4-1}

\usepackage{amsmath}
\usepackage{amssymb}
\usepackage{amsxtra,color}
\usepackage{latexsym}
\usepackage{graphicx}
\usepackage[utf8]{inputenc}
\usepackage{MnSymbol}	

\usepackage{hyperref}

\definecolor{MyDarkGreen}{rgb}{0,0.6,0}
\definecolor{MyDarkBlue}{rgb}{0,0,0.8}
\definecolor{MyDarkRed}{rgb}{0.6,0,0.3}

\hypersetup{breaklinks=true, colorlinks=true,plainpages=true, linktocpage=true, linkcolor=MyDarkBlue, citecolor=MyDarkGreen, urlcolor=MyDarkRed, pdfborder={0 0 0},%
pdfauthor={},%
pdfsubject={Research article},
pdftitle={}%
}

\newcommand{\be}{\begin{equation}}
\newcommand{\ee}{\end{equation}}
\newcommand{\bra}[1]{\langle  #1 |}
\newcommand{\ket}[1]{| #1 \rangle}
\newcommand{\braket}[2]{\langle \: #1 \: | \: #2 \: \rangle}

\newcommand{\braOketTight}[3]{\langle  #1  |  #2 |  #3  \rangle}
\newcommand{\braOketRed}[3]{\langle  #1  \|  #2 \|  #3  \rangle}

\makeatletter
\newcommand*{\rom}[1]{\expandafter\@slowromancap\romannumeral #1@}
\makeatother
\newcommand{\kbf}{\mathbf{k}}
\newcommand{\khat}{\mathbf{\hat k}}
\newcommand{\tauXGD}{\tau_\mathrm{X}^\mathrm{(GD)}}

\newcommand{\EX}{E_\mathrm{X}}

\begin{document}

\title{Attosecond pulse characterization with coherent Rydberg wavepackets}

\author{Stefan \surname{Pabst}}
\email{stefan.pabst@cfa.harvard.edu}
\affiliation{ITAMP, Harvard-Smithsonian Center for Astrophysics, 60 Garden Street, Cambridge, MA 02138, USA}
\affiliation{Physics Department, Harvard University, 17 Oxford Street, Cambridge, Massachusetts 02138, USA}



\author{Jan~Marcus \surname{Dahlstr\"om}}
\email{marcus.dahlstrom@fysik.su.se}
\affiliation{ITAMP, Harvard-Smithsonian Center for Astrophysics, 60 Garden Street, Cambridge, MA 02138, USA}
\affiliation{Department of Physics, Stockholm University, AlbaNova University Center, SE-106 91 Stockholm}


\begin{abstract}
We propose a new technique to fully characterize the temporal structure of extreme ultraviolet pulses by ionizing a bound coherent electronic wavepacket.
The populated energy levels make it possible to interfere different spectral components leading to quantum beats in the photoelectron spectrum as a function of the delay between ionization and initiation of the wavepacket.
The influence of the dipole phase, which is the main obstacle for state-of-the-art pulse characterization schemes, can be eliminated by angle integration of the photoelectron spectrum.
We show that particularly atomic Rydberg wavepackets are ideal and that wavepackets involving multiple electronic states provide redundant information which can be used to cross-check the consistency of the phase reconstruction.
\end{abstract}

\maketitle

\section{Introduction}
Laser-based ultrafast optics allows for pulse durations on the time scale of a few femtoseconds, 
limited by the fundamental period of the optical light, which is shorter than the response time of any photodetector. Ingenious techniques, based on nonlinear optics, have been developed to characterize the temporal structure of these laser pulses~\cite{Walmsley:09}. 
In 2001 the so-called femtosecond barrier was broken by generating coherent pulses in the (extreme) ultraviolet (UV) range by high-order harmonic generation \cite{LewensteinPRA1994}, in the form of isolated pulses \cite{HentschelNature2001,ZhZh-OptLett-2012} and pulse trains \cite{PaulScience2001}, which signaled the start a new era of ultra-fast experiments, known as attosecond physics \cite{KrIv-RMP-2009}. 

Recent experiments measuring the ionization delay between different orbitals of condensed matter system~\cite{CavalieriNature2007} and atoms~\cite{SchultzeScience2010,KlunderPRL2011} on the attosecond time scale have evidenced the need for full knowledge (amplitude + phase information) about the ionizing attosecond pulse when studying electron motion. 
However, the characterization of broad-band attosecond pulses remains challenging, because  techniques for nonlinear optics cannot be directly extended into the UV regime. This is due to  two main reasons: (i) low photon fluxes associated with the UV pulses do not favor nonlinear processes and (ii) high photon energies generally leads to ionization of the nonlinear medium, which implies target-dependent complex-valued susceptibilities that obscure the field-reconstruction procedure. 

For attosecond pulses, all current characterization techniques rely on ionizing an electron and dressing the photoelectron with a delayed near-infrared (NIR) pulse, which modulates the momentum of the electron~\cite{ChiniNaturePhotonics2014}.
This is the principle of the commonly used frequency-resolved optical gating for complete reconstruction of attosecond bursts (FROG-CRAB) method \cite{MairessePRA2005}. 
Even though the name of the technique contains attosecond, which implies broad spectral bandwidths, it relies on a narrow-band approximation to be able to apply the FROG retrieval algorithm~\cite{GaGo-AppPhysB-2008,WaCh-JPB-2009}.
An alternative to FROG-CRAB is the phase retrieval by omega oscillation filtering (PROOF) method \cite{ChiniOE2010}, where the IR field that dresses the continuum is chosen to be weak.

In both methods, several approximations are assumed. 
Most importantly, the dipole phases of the bound-continuum and continuum-continuum transitions are neglected without justification. 
Today it is known, from several relative ionization delay measurements, that these phases can lead to delays of several tens of attoseconds and can have non-trivial energy dependencies~\cite{CavalieriNature2007,SchultzeScience2010,KlunderPRL2011,GuenotJPB2014,PalatchiJPB2014}. 
In practice, these dipole phase effects translate directly into incorrectly reconstructed spectral phases of the attosecond pulse, and therefore, to the an uncertaintly in the retrived  pulse shape.  
Corrections to FROG-CRAB and PROOF methods have been proposed \cite{ZhangPRA2010,NagelePRA2011,DahlstromJPB2012,LaurentOE2013}, and, for noble gas atoms, detailed target-dependent response times have been computed by numerical methods that take into account both electron correlation and laser-driven dynamics~\cite{MoorePRA2011,PazourekPRL2012,DahlstromPRA2012,FeistPRA2014}. 
Still, uncertainties remain regarding the discrepancy between experiment and theory.
A new method that could measure the properties of the attosecond pulses independent of the complex atomic response would improve the accuracy of the attosecond pulse characterization.

We propose a new method of measuring the spectral phase of the attosecond pulse by ionizing a bound electronic wavepacket by the attosecond pulse and measuring the photoelectron spectrum.
The key differences to FROG-CRAB and PROOF are that: 
(i) the intermediate states are {\it bound}, 
(ii) the  pump and probe steps are {\it sequential} and  
(iii) the photoelectron measurement is integrated over all detection angles.   
As we will demonstrate, these three points result in an elimination of the dependence on dipole phases.  

The details of the preparation of the bound electronic wavepacket (pump) are not important,
but the pump process must be completed before the attosecond pulse starts the ionization  proceess (probe), so that pump and probe steps are sequential. 
We mention that Rydberg states have been exploited before in attosecond experiments, but with the focus on characterizing the bound electronic wavepacket and not to characterize UV pulses~\cite{MaRe-PRL105-2010,PaSc-PRA-2016}. In any case, all proposed methods have in common that they depend intrinsically on the dipole phases. 

We will show that this dipole phase-dependence of the probe step is eliminated if the photoelectron detection is integrated over all angles provied that  
the parent cation is spherically symmetric, and that correlation effects can be ignored. 
We propose to use atomic alkali metals because the spherical symmetry condition is perfectly fulfilled and the preparation of the Rydberg wavepacket can be performed using optical laser pulses (pump) rather than UV pulses. The influence of correlation can be reduced to a minimum by considering a highly excited Rydberg wavepacket. 

In addition, our method provides an intrinsic self-consistency check for the retrieved phases when  the electronic wavepacket consists of more than two states. To our knowledge, no other ultrafast pulse reconstruction method (for femtosecond or attosecond pulses) possess such a powerful feature. Multi-level wavepackets are also beneficial when analyzing pulses with more complex spectral structure, such as in pulse trains, where certain spectral regions have vanishing intensity. Being able to choose relatively freely the energy distance between states, which is especially true for Rydberg states, phase dependency within {\it and} across harmonics can be probed. Our method is not limited to measurements on the attosecond time scale, but it can also be applied to longer pulses in the UV or optical range. The sequential pump-probe setup with a variable amount of spectral shearing makes the propose method ideal for evaluation of seeded FEL pulses~\cite{MuNi-NatPhoton-2012,GaRi-PRL-2015}, where the FEL pulse (probe) is separately produced by a phase-locked replica of the laser pulse (pump).   

In Sec.~\ref{sec:theory}, we discuss the theory behind our proposed method.
In Sec.~\ref{sec:results.pulse}, we show examples how to reconstruct the spectral phase of an isolated attosecond pulse and an attosecond pulse train by using a Rydberg wavepacket in potassium.
The influence of electronic correlation effects are discussed in Sec.~\ref{sec:results.estruc} as well as the possibility to extract the dipole phase of the bound-continuum transition when analyzing the directional photoelectron spectrum as well.
We show that the dipole phase dependence due to inner-shell correlations and Fano resonances is much smaller in the angle-integrated photoelectron spectrum than in the directional photoelectron spectrum.

Atomic units are used: $\hbar=|e|=4\pi\epsilon_0=1$ unless otherwise stated.

\section{Theory}
\label{sec:theory}
\begin{figure}[t!]
  \includegraphics[width=\linewidth]{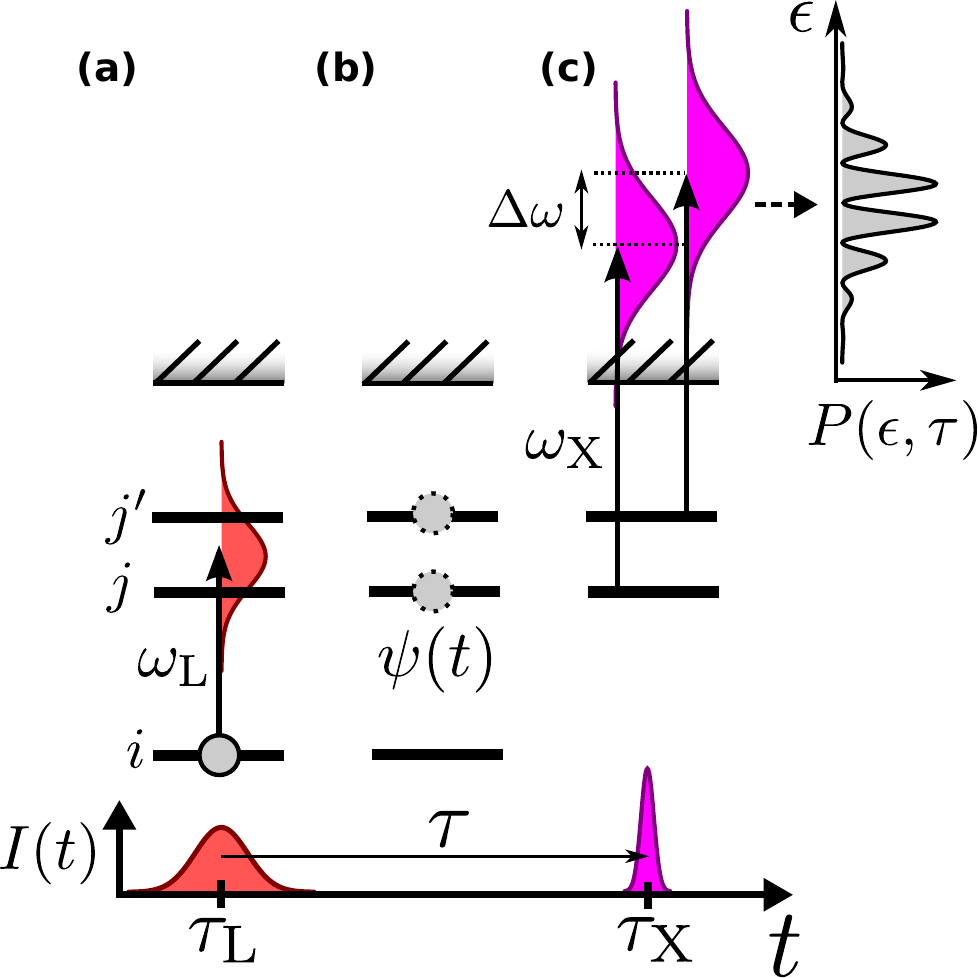}
  \caption{Sketch of the method to characterize an attosecond laser pulse  using a bound electron wavepacket.
  (a) Preparation of the wavepacket with laser pulse.
  (b) Field-free propagation of the electronic wavepacket $\psi$ for the duration $\tau$.
  (c) Ionization of the wavepacket by an attosecond UV pulse.
  Due to the coherent superposition of different electronic states, different spectral components of the UV pulse are absorbed to reach the same final continuum state with energy $\epsilon$.
  The interferences between the different spectral components contain the spectral phase information of the UV pulse, which can be extracted by varying the delay $\tau$.
  }
  \label{fig:overview}
\end{figure}

In this paper we propose a novel approach for UV pulse characterization that relies on a coherent superposition of electronic states in the target system to interfere different spectral components of the UV pulse with each other.
The basic idea is illustrated in Fig.~\ref{fig:overview}.
The first step is the preparation of a coherent superposition of electronic states,
\begin{align}
  \label{eq:wavepacket}
  \ket{\psi(t)}
  &=
  \sum_j c_j\,e^{-i\epsilon_j t} \, \ket{j}
  ,
\end{align}
where $\epsilon_j$ is the energy of state $j$. 
The amplitudes $c_j$ depend highly on the preparation process.
The superposition can be achieved in various ways and the specific method does not matter here.
An obvious one is one- or multi-photon absorption with a broad band pulse.
Another one could be a scheme based on Rabi oscillations.
For characterizing the pulse it is important that the excited states have the same parity, or 
more specifically, the electron in the intermediate states must be able to reach the final state via one-photon absorption. However, having states with opposite parity within the wavepacket is not harmful as they do not contribute to the interference. 

Step 2 is field-free propagation of the electronic wavepacket until time $\tau$ where, in step 3, the UV pulse of interest,
\begin{align}
  \label{eq:pulse-uv}
  E_X(\omega,\tau)
  &=
  |E_X(\omega)| \,
  e^{i\omega\tau+i\phi_X(\omega)}
  ,
\end{align}
ionizes the wavepacket (see Fig.~\ref{fig:overview}).
$|E_X(\omega)|$ and $\phi_X(\omega)$ are the amplitude and phase of the unshifted UV pulse, namely for $\tau=0$.
In the time domain the electric field is given by
\begin{align}
\tilde E_X(t,\tau) 
&=
\frac{1}{2\pi}\int \!\! d\omega \ E_X(\omega) \, e^{-i\omega (t-\tau)+i\phi_X(\omega)}
=
\tilde E_X(t-\tau,0),
\end{align}
where we show that a pulse with a given $\tau$ on the left side corresponds to a pulse delayed by $\tau$ of the right side. 
As step 3 is a simple one-photon ionization process, the photoelectron amplitude is given by first-order perturbation theory,
\begin{align}
\label{eq:pes_amp}
c_f(\tau)
&=
\lim_{t'\rightarrow\infty}\frac{1}{i} 
\sum_j  c_j \, d_{fj} 
\int^{t'} \!\!\! dt \ \tilde E_X(t,\tau) \, e^{i\omega_{fj}t}
\nonumber \\
&=
-i\sum_j  c_j \, d_{fj} \, |\EX(\omega_{fj})| \, e^{i\omega_{fj}\tau+i\phi_X(\omega_{fj})}
,
\end{align}
where $d_{fj}$ is the dipole transition moment from the bound state, $j$, to the continuum state, $f$.
The lower bound of the time integral can be set to $-\infty$ because we assume that the ionization step of the wavepacket is truly sequential, namely that the UV pulse ionizes the wavepacket after the wavepacket has been prepared so that the $c_j$ amplitudes can be treated as constants.   

Different UV energies, $\omega_{fj}=\epsilon_f-\epsilon_j$, are required to reach the same final state, $\ket{f}$, from different bound states of wavepacket, $\ket{j}$ [see Fig.~\ref{fig:overview}(c)]. 
The interference of the different ionization pathways make it possible to learn more about the phase relation between the different spectral components of the UV pulse.
The final photoelectron spectrum reads
\begin{align}
\label{eq:pes_directional}
P({\bf k}_f,\tau)
&=
|c_f(\tau)|^2 
=
\sum_{j} \left( A_{j} + \sum_{j'>j} B_{jj'}\,\cos \Theta_{jj'} \right)
,
\end{align}
where ${\bf k}_f$ is the momentum of the photoelectron with energy $\epsilon_f=k_f/2$.
The variable $A_{j}$ is the isolated contribution from the wavepacket state $\ket{j}$ to $\ket{f}$, 
while $B_{jj'}$ and $\Theta_{jj'}$ are the amplitude and phase of quantum interference between different parts of the wavepacket with energies $\epsilon_{j'}>\epsilon_{j}$.  The explicit form of the coefficients in Eq.~(\ref{eq:pes_amp}) are
\begin{subequations}
\label{eq:para-directional}
\begin{align}
\label{eq:para-directional.A}
A_{j}
&=
|c_j|^2 \, |d_{fj}|^2\, |\EX(\omega_{fj})|^2,  
\\
\label{eq:para-directional.B}
B_{jj'}
&=
2 \left|c_j \, c_{j'}\right| \  
\left| d_{fj} \, d_{fj'} \right| \ 
\left| \EX(\omega_{fj}) \EX(\omega_{fj'}) \right|
\\
\label{eq:para-directional.Theta}
\Theta_{jj'}
&=
\mathrm{arg}\big[
c_j\,c_{j'}^* \ d_{fj}\, d_{fj'}^* \  \EX(\omega_{fj},\tau) \, \EX^*(\omega_{fj'},\tau)
\big]
.
\end{align}
\end{subequations}
The phase of the interference pattern can be rewritten to clearly show the quantum beating of the probability, at each energy difference $\omega_{j'j}=\omega_{fj}-\omega_{fj'}>0$, as   
\begin{align}
\label{eq:Theta-directional}
\Theta_{jj'}
&=
\omega_{j'j}\, \tau
+
\phi_X^{(jj')}(\epsilon_f)
+
\phi_D^{(jj')}(\epsilon_f)
+
\phi_I^{(jj')}
\end{align}
where the phase differences read
\begin{subequations}
\label{eq:taus}
\begin{align}
\label{eq:taus.pulse}
\phi_X^{(jj')}(\epsilon_f)
&=
\phi_\mathrm{X}(\omega_{fj})-\phi_\mathrm{X}(\omega_{fj'})
\\\nonumber
&\approx \tauXGD( \bar\omega_{jj'}^{(f)} ) \ \omega_{j'j} 
,
\\
\label{eq:taus.dipole}
\phi_D^{(jj')}(\epsilon_f)
&=
\mathrm{arg}[d_{fj}]-\mathrm{arg}[d_{fj'}]
,
\\
\label{eq:taus.wavepckt}
\phi_I^{(jj')}
&=
\mathrm{arg}[c_{j}]-\mathrm{arg}[c_{j'}]
.
\end{align}
\end{subequations}
Equation~(\ref{eq:taus.pulse}) defines the group delay, $\tauXGD(\bar\omega_{jj'}^{(f)})$, of the UV pulse 
computed as a finite-difference derivative at the mean photon energy $\bar\omega_{jj'}^{(f)}=(\omega_{fj}+\omega_{fj'})/2$.
The frequency dependence of the group delay is mapped onto the photoelectron spectrum as energy dependent phase differences.
$\phi_I^{(jj')}$ is inherent to the preparation process of the bound wavepacket and independent of the photoelectron energy.
In contrast,  $\phi^{(jj')}_X(\epsilon_f)$ and $\phi^{(jj')}_D(\epsilon_f)$ are functions of the photoelectron energy, $\epsilon_f$.
Note that the dipole transition elements, $d_{fj}$, are generally complex valued leading to nontrivial energy dependence in the phase differences, $\phi^{(jj')}_D(\epsilon_f)$. 

The main goal in characterizing the UV pulse is the determination of $\phi_X^{(jj')}(\epsilon_f)$. If this phase difference is measured accurately with a suitably small $\omega_{j'j}$, the spectral phase $\phi_X(\omega)$ can be fully reconstructed (up to an absolute value) by numerical integration of the phase difference.   
The oscillations in the photoelectron spectrum depend on $\phi_D^{(jj')}(\epsilon_f)$ and $\phi_I^{(jj')}(\epsilon_f)$.
The intrinsic phase shift $\phi_I^{(jj')}$ does not present an obstacle as it is energy independent and corresponds to a constant shift of the pump-probe delay by $\phi_I^{(jj')}/\omega_{j'j}$.
The energy dependence of the ionization step, $\phi_D^{(jj')}(\epsilon_f)$, is, in general, an obstacle and prevents the determination of $\phi_X^{(jj')}(\epsilon_f)$ as a function of photoelectron energy.
 
For spherically symmetric systems, and more specifically for spherically symmetric cations without correlation effects, the radial dipole moment to any energy eigenstate (continuum or bound) can be chosen to be real, c.f. Ref.~\cite{Friedrich-AMO-book}. 
As a result, the angle-integrated photoelectron spectrum does not have an energy-dependent $\phi_D^{(jj')}$-phase anymore. 
To see this, we start with a partial wave expansion of the final continuum state,
\begin{align}
\label{eq:outgoing-wave}
\psi^-_{\mathbf{k}}(\mathbf{r})
&=
\frac{1}{k^{1/2}}\sum_{L=0}^\infty\sum_{M=-L}^L i^L e^{-i\eta_L} Y_{LM}^*(\khat) \, Y_{LM}(\mathbf{\hat r}) \, R_{\epsilon L}(r)
,
\end{align}
where $\kbf=k\khat$ is the momentum of the photoelectron and its energy is given by $\epsilon_f=k^2/2$~\cite{St-Springer-1980}.
Note that $\psi^-_{\mathbf{k}}$ is momentum normalized, 
$\braket{\psi^-_{\mathbf{k}}}{\psi^-_{\mathbf{k'}}}=\delta(\mathbf{k}-\mathbf{k'})$, 
while the real radial wavefunctions are chosen to be energy normalized, 
$\braket{R_{\epsilon L}}{R_{\epsilon' L}}=\delta(\epsilon-\epsilon')$. 
Inserting Eq.~\eqref{eq:outgoing-wave} as the final state in Eq.~\eqref{eq:pes_directional} leads to
\begin{align}
\label{eq:pes_angmom}
P(\mathbf{k},\tau)
=&\
|c_\mathbf{k}(\tau)|^2
\\ \nonumber
=&\
\frac{1}{k}\sum_{j}\sum_{j'}c_j c_{j'}^* \sum_{L,M} \sum_{L',M'} 
i^{L-L'} \,
Y_{L'M'}(\khat) \, Y^*_{LM}(\khat)  
\\ \nonumber
&\times
e^{-i\eta_L+i\eta_{L'}} \
d_{f j}\,d^*_{{f'} j'} \ \EX(\omega_{fj},\tau)\,\EX^*(\omega_{fj'},\tau )
,
\end{align}
where we used the partial-wave basis $\ket{f}=\ket{\epsilon_f L M}$ and $d_{f j}=\braOketTight{f}{\hat d}{j}=\braOketRed{\epsilon_f L}{r}{n_jL_j} \ C^{LM}_{L_jM_j ; 1 \mu} \sqrt{2L+1}^{-1}$ with $\mu$ being the polarization of the ionizing UV pulse.
In Eq.~(\ref{eq:pes_angmom}) the sums on $j$ and $j'$ run over all excited states of the prepared wavepacket.  
Since the radial part of the continuum states and the bound state are chosen to be real functions, the reduced matrix element, $\braOketRed{\epsilon_f L}{r}{n_jL_j}$, and the dipole, $d_{f j}$, are real and do not contribute to a phase difference in the photoelectron spectrum. 

Equation~\eqref{eq:pes_angmom} depends on the scattering phase differences between all final partial waves states resulting in non-trivial angle and energy dependencies.
By performing an angle-integrated measurement, the dependencies on the dipole phase can be eliminated,
\begin{align}
\label{eq:pes_angle-int_explicit}
P(\epsilon,\tau)
&=
\int \!\! d\Omega_\kbf \ |c_\kbf(\tau)|^2
\\ \nonumber 
&=\sum_{L_f,j',j} c_{j} c_{j'}^* d_{f j} d_{f j'} 
\ \EX(\omega_{fj},\tau)\EX^*(\omega_{fj'},\tau)
,
\end{align} 
where we have used the orthogonality of the spherical harmonics to reduce the sums to one sum over final angular momentum, $L_f$, than can be reached by one photon from the prepared wavepacket with $L_j$. 
Rewriting Eq.~\eqref{eq:pes_angle-int_explicit} by grouping contributions form the same ($j=j'$) and different electronic states ($j\neq j'$) together, we obtain
\begin{align}
 \label{eq:pes_angle-int}
 P(\epsilon,\tau)
 &=
 \sum_j \left( \bar A_{j} + \sum_{j'>j} \bar B_{jj'} \cos\bar\Theta_{jj'} \right)
\end{align} 
where the variables read
\begin{subequations}
\label{eq:para-int}
\begin{align}
\label{eq:para-int.A}
\bar A_j
&=
\sum_{L_f} |c_j|^2 \ |d_{f j}|^2 \ |\EX(\omega_{fj})|^2/k
,
\\
\label{eq:para-int.B}
\bar B_{jj'}
,
&=
\sum_{L_f}
d_{f j} d_{f j'}   \
\left| c_j c_{j'} \right| \
\left| \EX(\omega_{fj}) \, \EX(\omega_{fj'}) \right|/k
,
\\
\label{eq:para-int.theta}
\bar\Theta_{jj'}
&=
\omega_{j'j}\,\tau
+
\phi_X^{(jj')}(\epsilon_f)
+
\phi_I^{(jj')}
,
\end{align}
\end{subequations}
where we write the index $f$ explicitly on  $L_f$ for clarity. 
The phase term, $\Theta_{jj'}$, is independent of atomic scattering (or dipole) phases and of $L_f$. 
As the dipoles can be positive or negative, there should be no absolute magnitude around $d_{f j} d_{f j'}$ in Eq.~\eqref{eq:para-int.B}.
Consequently, the phase modulation depends on the properties of the wavepacket preparation, $\phi_I^{(jj')}$ and on the phases of the UV field, $\phi_X^{(jj')}(\epsilon_f)$, but not on the final scattering phases $\eta_{L_f}(k)$  --- exactly what we wanted. 

When more than two states are involved, say $N$, in the electronic wavepacket, $\psi(t)$, we obtain several phase differences, $\phi_X^{(j_1j_2)}$ in $\binom{N}{2}$ different $\omega_{jj'}$ oscillations.
It turns out there exist groups of three phase differences that obey the relation
\begin{align}
  \label{eq:phiX-relations}
  \phi_X^{(j_1j_3)}(\epsilon_f)  
  &=
  \phi_X(\omega_{fj_1})
  -
  \phi_X(\omega_{fj_3})
  \\\nonumber 
  &=
  \phi_X(\omega_{fj_1})
  -
  \phi_X(\omega_{fj_2})
  +
  \phi_X(\omega_{fj_2})
  -
  \phi_X(\omega_{fj_3})
  \\\nonumber
  &=
  \phi_X^{(j_1j_2)}(\epsilon_f)
  +
  \phi_X^{(j_2j_3)}(\epsilon_f)
  .
\end{align}
This connection can be used to cross-reference the extracted phases and to check the consistency of the retrieved phases. 
Eq.~\eqref{eq:phiX-relations} holds also true for the retrieved phase differences up to a constant phase difference, because $\phi_I^{j_1j_2}$ does not need to fulfill any special relation. 
However, $\phi_I^{j_1j_2}$ can be normally  determined for most common preparation methods, e.g. by applying femtosecond pulse characterization to the laser pulses that are used to prepare the bound electron wavepacket.

Our scheme to characterize UV pulses has several advantages that can be summarized as follows:
\begin{itemize}
  \item Ionization pathways that lead to different final ionic states or different angular momenta of the photoelectron do not interfere, and consequently, do not affect the phase reconstruction.
  \item Using Rydberg states makes it possible to probe spectral phase differences bridging arbitrarily small energy distances as the energy difference between Rydberg states goes like $n^{-3}$. The phase difference, $\phi_X^{(n+1,n)}(\omega)$, approaches the exact phase derivative times the frequency difference, $\partial_w \phi_X(\omega)\ \omega_{n+1,n}$.  
  \item Having small {\it and} large energy differences, $\omega_{j'j}$, is beneficial to probe simultaneously phase differences between close-by and distance spectral phase components. Especially for pulse trains this is a convenient feature as the spectral phase within and between harmonics can be measured. 
  \item There is no constrain on the polarization of the UV pulse. It works for linearly, circularly, and elliptically polarized pulses.
  \item Involving $N$ electronic states (with the same angular momentum) in the wavepacket leads to $\binom{N}{2}$ beating frequencies in the photoelectron spectrum providing redundant information [see Eq.~\eqref{eq:phiX-relations}] to cross-check the consistency of the data.
\end{itemize}

In theory, electron correlation result in complex-valued dipole moments, which may affect the pulse reconstruction, even if photoelectron is detected over all angles. As we will show in Sec.~\ref{sec:results.estruc} for the case of Rydberg wavepackets, these effects are surprisingly small even in strongly correlated energy regions close to autoionizing resonances.


\section{Results}
\label{sec:results}

In this section, we first use the Rydberg wavepacket to characterize a single attosecond pulse (see Secs.~\ref{sec:results.pulse.single+2level} and \ref{sec:results.pulse.single+multilevel}) and an attosecond pulse train (see Sec.~\ref{sec:results.pulse.train+multilevel}).
We choose potassium as our system of interest. 
Atomic alkali metals possess several advantages:
(1) After removing the the outermost $s$ electron, alkali cations are closed-shell system with relatively small amounts of correlation. 
The mean-field potential of alkali cations is spherically symmetric fulfilling exactly the conditions for real reduced dipole matrix elements.
(2) The excitation energies of the Rydberg states in all alkali metals lie within 1.4~eV and 5.4~eV~\cite{NIST_website}. 
This is a very convenient energy range as it is easily accessible with common laser systems and standard nonlinear optics.

%
%

In Sec.~\ref{sec:results.estruc}, we use photoionization of the Rydberg wavepacket to study the electronic structure properties.
Avoiding the angle integration, the oscillations in the photoelectron spectrum contain also information about the dipole phases (see Sec.~\ref{sec:results.estruc.directional}).

Electronic correlations result in complex-valued reduced dipole moments and, consequently, the oscillations in the photoelectron spectra depend on the dipole phases. 
We study the influence of electronic correlation in the case of potassium (see Sec.~\ref{sec:results.estruc.directional}) and in neon near the $2s^{-1}3s$ Fano resonance (see Sec.~\ref{sec:results.estruc.fano}).

All results obtained by explicit time propagation~\cite{GrSa-PRA-2010} utilize the {\sc xcid} program~\footnote{S. Pabst, L. Greenman, A. Karamatskou, Y.-J. Chen, A. Sytcheva, O. Geffert, R. Santra--\textsc{xcid} program package for multichannel ionization dynamics, DESY, Hamburg, Germany, 2015, Rev. 1790} with the following numerical parameters~\footnote{
A pseudo-spectral grid with a radial box size of 120~$a_0$, 750 grid points, and a mapping parameter of $\zeta=0.5$ are used. 
There is no complex absorbing potential.
The splitting function starts around $R_\text{split}=70~a_0$, a smoothness of 2~$a_0$, and splitting interval of $dt_\text{split}=2$~a.u. 
The maximum angular momentum is 2 and Hartree-Fock orbitals up to an energy of 5~$E_h$ are considered. 
The propagation method is Runge-Kutta 4 with a time step $dt=0.05$~a.u.
}
and the wavefunction splitting method to analyze the photoelectron spectrum as described in Ref.~\cite{KaPa-PRA-2014}.
This method has been successfully used to analyze photoelectron spectra in the multiphoton regime~\cite{MaKa-NatComm-2015,PaWa-PRA-2015}.
In this work we are staying in the one-photon regime. 
Correlation effects for K are studies by means of the random phase approximation with exchange (RPAE) by adapting existing numerical codes 
\cite{DahlstromJPB2014} with the initial state being a given $np$ virtual orbital obtained by the Hartree-Fock equation of the $K ^+$ core, as described in text books on many-body perturbation theory, c.f.~\cite{mbpt,Amusia1990}. 
TDCIS is also used for studying the correlation effects in neon, as it has been shown to describe well many-body physics in the attosecond~\cite{PaSa-PRL-2011} and strong-field regime~\cite{PaSa-PRL-2013}.

\subsection{Pulse characterization}
\label{sec:results.pulse}

\subsubsection{Single Pulse + 2-level wavepacket}
\label{sec:results.pulse.single+2level}

As the Rydberg wavepacket can be easily prepared with common laser systems, we omit the description of its preparation. In our first example, we choose the wavepacket in potassium to be a coherent superposition between the $3p^64p$ and $3p^65p$ Rydberg states,
\begin{align}
  \label{eq:k.4p5p-wavepacket}
  \ket{\Psi(t)}
  &=
  \frac{1}{\sqrt{2}}
  \sum_{n=4,5}
    e^{-i\,\varepsilon_{np}\,t}
    \,
    \ket{np}
  .
\end{align}
We omitted the reference to the inner-shell configuration as we are only interested in the Rydberg electron and the ionic configuration will not change (see Sec.~\ref{sec:results.estruc.fano} for a more general discussion and the influence of interchannel coupling). 
Potassium is treated within the HF level, where the HF procedure is done for $K^+$.
The $4s$ electron, which gets excited feels, therefore, the mean-field potential of $K^+$.
Interchannel are ignored and intrachannel correction are not needed as the HF procedure is done for $K^+$ and not for neutral $K$.

In the probe step, the $4p$--$5p$ Rydberg wavepacket is ionized by the attosecond pulse, which we want to characterize.
We consider an isolated attosecond pulse with constant spectral phase ($\phi_X(\omega)=0$), with a quadratic spectral phase ($\phi_X(\omega)=100/E^2_h~(\omega-\omega_0)^2$), and with a cubic spectral phase ($\phi_X(\omega)=100/E^3_h~(\omega-\omega_0)^3$). 
Each attosecond pulse has the central photon energy $\omega_0=74$~eV ($=2.72$~a.u.) and a full-width-half-maximum (FWHM) spectral width (of the intensity profile) of 7.4~eV ($=0.27$~a.u.) corresponding to a pulse duration of 247~as for a Fourier-limited pulse. 
The pulse with linear chirp (quadratic phase) is symmetrically stretched in time increasing the pulse duration, while the pulse with quadratic chirp (third-order phase) experiences an asymmetric pulse deformation.

\begin{figure}[t!]
  \includegraphics[width=\linewidth]{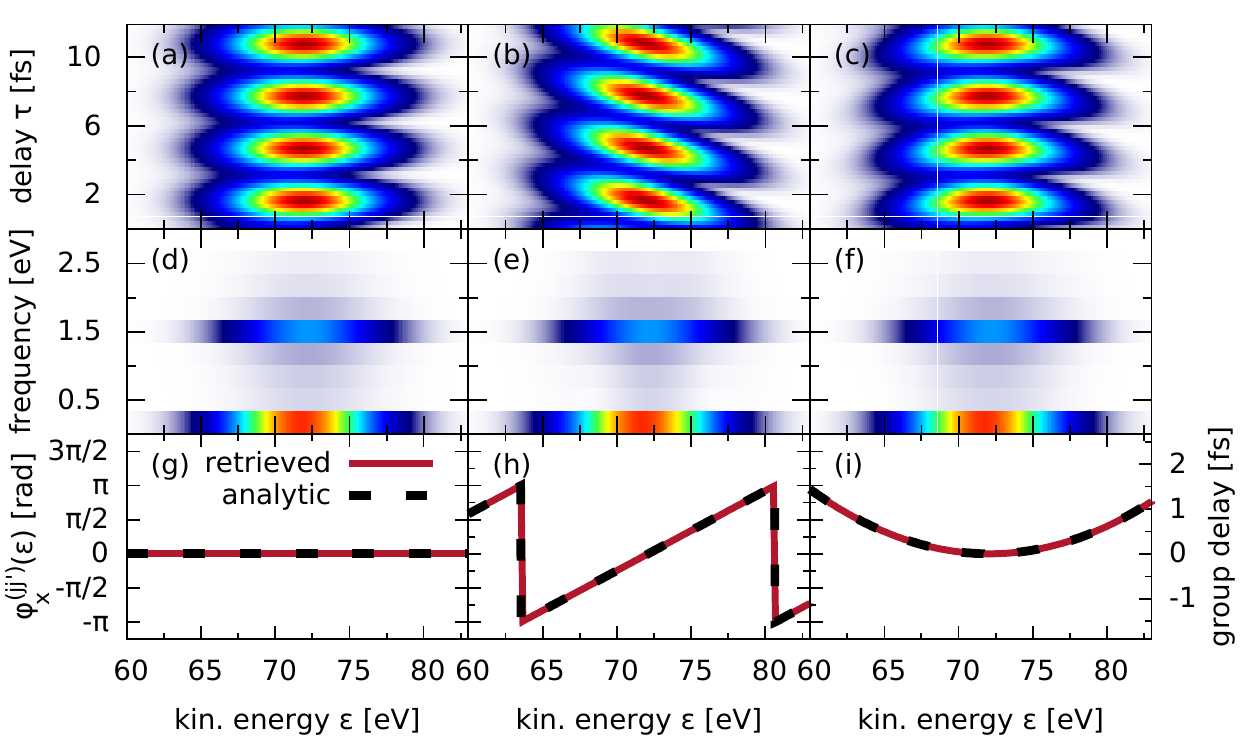}
  \caption{
  (a-c) Photoelectron spectra of the $4p$--$5p$ Rydberg wavepacket ionized by an attosecond pulse with (a) no chirp, (b) a linear chirp, and (c) a quadratic chirp ionizing as a function of the pulse delay.
  (d-f) Photoelectron spectra where the delay axis is Fourier transformed. The spectrum contains a static signal and an oscillating component with frequency 1.36~eV.
  (g-i) Retrieved phase differences of the 1.36~eV component. 
  Analytic results are shown as black dashed lines.
  The corresponding group delay of the phase differences is shown on the right.
  }
  \label{fig:k45p_1pulse}
\end{figure}

In Figs.~\ref{fig:k45p_1pulse}(a-c) the photoelectron spectra, $P(\epsilon,\tau)$, are shown for the attosecond pulse with (a) no chirp, (b) a linear chirp, and (c) a quadratic chirp.
The linear and quadratic chirps are visible in Figs.~\ref{fig:k45p_1pulse}(b) and (c) as linear and quadratic phase differences of the beating patterns, respectively. 
This is exactly the behavior we expect according to Eqs.~\eqref{eq:taus.pulse} and \eqref{eq:para-int.theta}, because the phase differences in the spectrum are related to the phase differences in the attosecond pulse.

In Figs.~\ref{fig:k45p_1pulse}(d-f) the photoelectron spectra, $P(\epsilon,\nu)$, are shown where the delay axis is Fourier transformed. 
In all three cases a static signal and an oscillating signal with the beating frequency corresponding to $\epsilon_{5p}-\epsilon_{4p}=1.36$~eV is visible. 
The discreteness in the data is defined by the the length of the delay range used in the calculations, which is 500~a.u. (12.1~fs).

The retrieved phases of the 1.36~eV beating signal for the different chirped attosecond pulses are shown in Figs.~\ref{fig:k45p_1pulse}(g-i).
According to Eq.~\eqref{eq:para-int.theta}, these phases differences correspond to $\phi_X^{(jj')}(\epsilon_f)+\phi_I^{(jj')}$, where $j=4p$ and $j'=5p$. 
The intrinsic phases of the Rydberg wavepacket $\phi_I^{(jj')}$ are subtracted in Figs.~\ref{fig:k45p_1pulse}(g-i) (see Sec.~\ref{sec:theory} why $\phi_I^{(jj')}$ can be ignored).
The analytic solutions of the phase differences, $\phi_X^{jj'}(\epsilon_{f})$, are plotted as black-dashed lines, and agree perfectly with the retrieved phases.


\subsubsection{Single Pulse + multi-level wavepacket}
\label{sec:results.pulse.single+multilevel}

To characterize a pulse, the Rydberg wavepacket does not need to be a superposition of only two Rydberg states.
Involving $N>2$ states (with the same angular momentum $l$) in the Rydberg wavepacket is even more beneficial as $\binom{N}{2}$ beating frequencies occur in the photoelectron spectra---all containing spectral information about the attosecond pulse.

\begin{figure}[t!]
  \includegraphics[width=\linewidth]{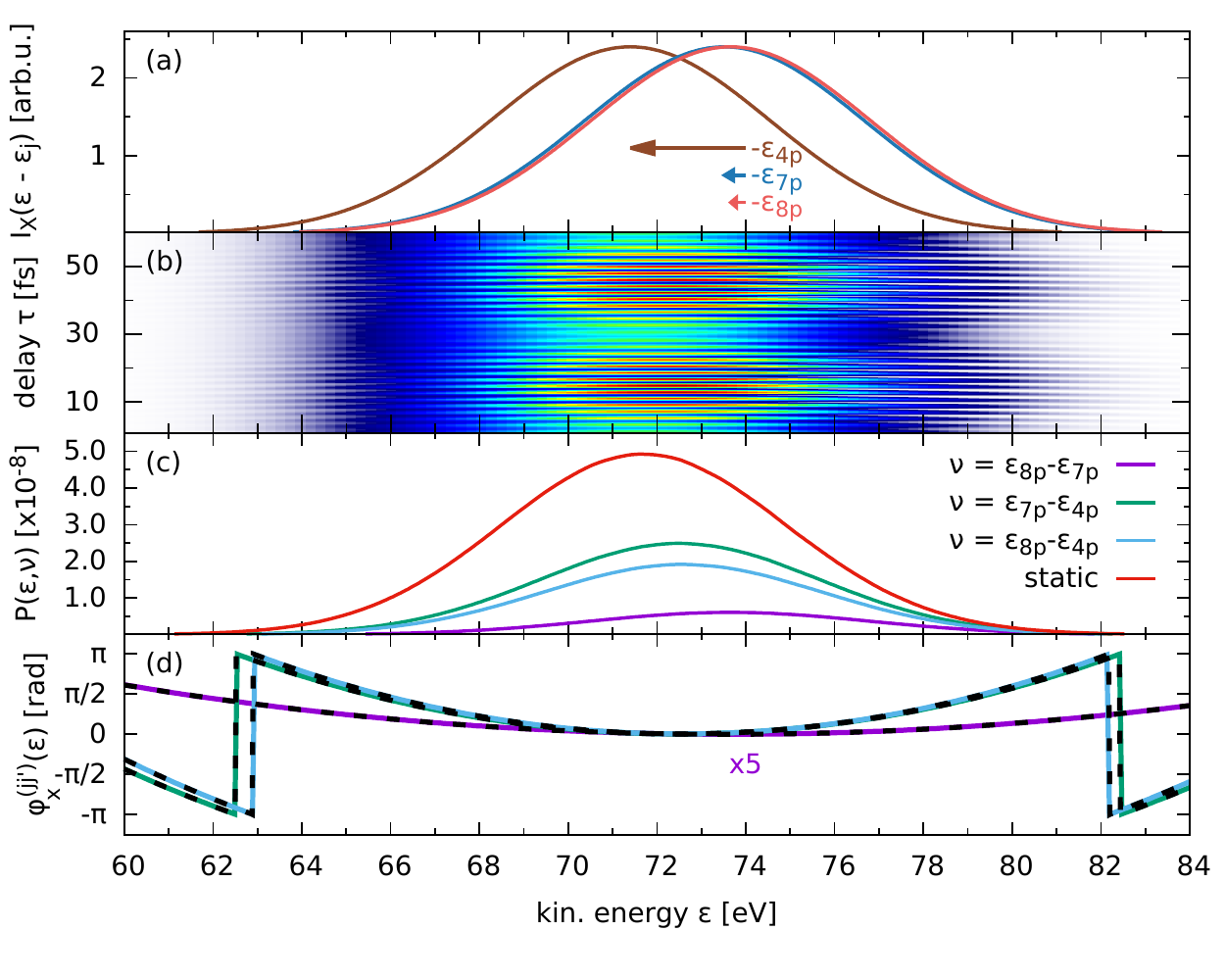}
  \caption{
  (a) Spectra of the attosecond pulse, $\EX(\omega_{fj})$, shifted by the Rydberg binding energies $\epsilon_{4p},\epsilon_{7p},$ and $\epsilon_{8p}$ (indicated by arrows).
  (b) Photoelectron spectrum of the $4p$--$7p$--$8p$ Rydberg wavepacket ionized by a quadratic-chirped single attosecond pulse.
  (c) The static and the 0.14~eV, 2.12~eV, and 2.26~eV oscillating signals of the photoelectron spectrum.
  (d) Retrieved phases of the three frequency components.
  The analytic results are shown as black dashed lines.
  }
  \label{fig:k478p_1pulse} 
\end{figure}

We consider the 3-level Rydberg wavepacket,
\begin{align}
  \label{eq:k.478p-wavepacket}
  \ket{\Psi(t)}
  &=
\frac{1}{\sqrt{3}}
  \sum_{n=4,7,8}
    e^{-i\,\varepsilon_{np}\,(t-t_0)} \,
    \ket{np}
  ,
\end{align}
which provides two high frequency beatings, $\omega_{84}=\epsilon_{8p}-\epsilon_{4p}=2.26$~eV and $\omega_{74}=\epsilon_{7p}-\epsilon_{4p}=2.12$~eV, and one low frequency beating $\omega_{87}=\epsilon_{8p}-\epsilon_{7p}=0.14$~eV.
The reference time, where all phases of the wavpacket are the same, is chosen to be $t_0=-600$~a.u.
In Fig.~\ref{fig:k478p_1pulse}(a), the pulse spectrum shifted by the ionization potentials towards lower energies, $\epsilon_j$, is shown indicating where the photoelectron spectrum will be located.

In Fig.~\ref{fig:k478p_1pulse}(b) the photoelectron spectrum, $P(\epsilon,\tau)$ when ionizing the 3-level Rydberg wavepacket with the attosecond pulse with cubic phase (quadratic chirp) of Sec.~\ref{sec:results.pulse.single+2level}.
The photoelectron spectra, $P(\epsilon,\nu)$, for the three frequencies $\nu=\omega_{47},\omega_{48},\omega_{78}$ are shown in Fig.~\ref{fig:k478p_1pulse}(c).

The retrieved phases and the analytic result match perfectly for the three beating frequencies [see Fig.~\ref{fig:k478p_1pulse}(d)].
All three curves are parabolas with curvatures $3\beta\,\omega_{j'j}$ depending on the quadratic chirp, $\beta=100/E^3_h$, and on the energy difference between the states, $\omega_{j'j}$.
Also the relations between $\phi_X^{(jj')}$ (see Eq.~\eqref{eq:phiX-relations}) hold for all combinations. 
This demonstrates the power of this method which provide an internal mechanism to confirm the consistency of the retrieved phases.

\subsubsection{Pulse train + multi-level wavepacket}
\label{sec:results.pulse.train+multilevel}

The three-level Rydberg wavepacket can be applied to probe the phase of attosecond pulse trains.
We choose a pulse train that consists of the 48\textsuperscript{th}, 49\textsuperscript{th}, and the 50\textsuperscript{th} harmonics of 800~nm. 
Each harmonic has a spectral (intensity) FWHM width of $\delta\omega=400$~meV. 
The 48\textsuperscript{th} harmonic has no spectral chirp but a phase difference of $\pi/4$ whereas the 49\textsuperscript{th} harmonic has a linear chirp, $\phi_X(\omega)=-0.5/\delta\omega^2 (\omega-\nu_{55})^2$, and the 50\textsuperscript{th} harmonic has a quadratic chirp, $\phi_X(\omega)=0.5/\delta\omega^3  (\omega-\nu_{56})^3 + \pi/3$.
Since the spectral spectrum of a pulse train is more structured than for an isolated pulse, the photoelectron spectrum contains more spectral features. 

%

\begin{figure}[t!]
  \includegraphics[width=\linewidth]{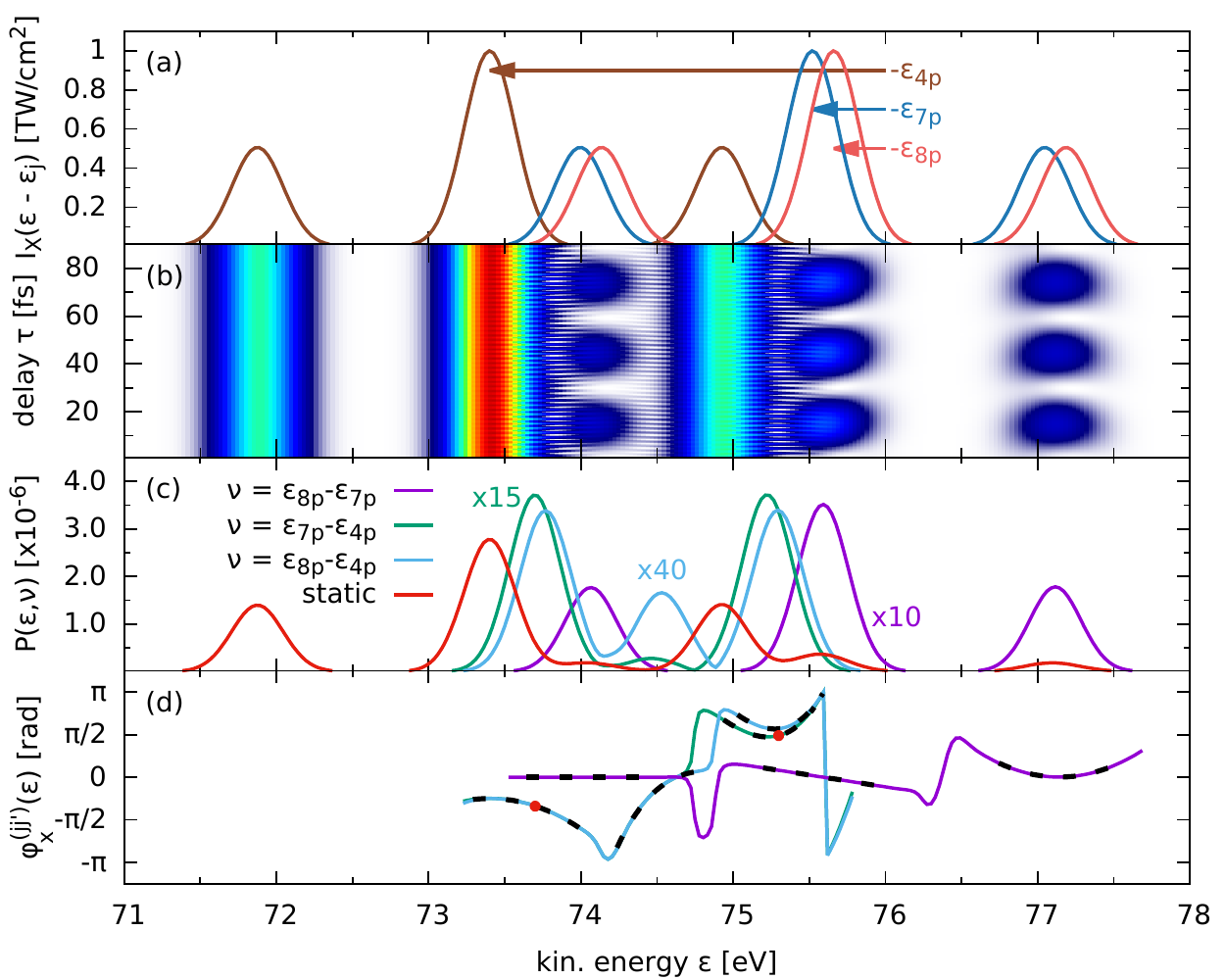}
  \caption{
  (a) Spectra of the attosecond pulse train, $\EX(\omega_{fj})$, shifted by the Rydberg binding energies $\epsilon_{4p},\epsilon_{7p},$ and $\epsilon_{8p}$ (indicated by arrows).
  (b) Photoelectron spectrum of the $4p$--$7p$--$8p$ Rydberg wavepacket ionized by the attosecond pulse train.
  (c) The static and the 0.14~eV, 2.12~eV, and 2.26~eV oscillating components of the photoelectron spectrum.
  (d) Retrieved phases of the three frequency components.
  The analytic results are shown as black dashed lines.
  The highlighted phases (red circles) at 73.7~eV and 75.3~eV for the $4p$--$7p$ beating are used to determined the relative phases between harmonics.
  }
  \label{fig:k478p_train} 
\end{figure}

By using the $4p$--$7p$--$8p$ Rydberg wavepacket [see Eq.~\eqref{eq:k.478p-wavepacket}] from Sec.~\ref{sec:results.pulse.single+multilevel},
we can interfere spectral components within the same harmonic and across harmonics.
The overlap of the pulse spectrum shifted by the binding energies $\epsilon_j$ visualized it in Fig.~\ref{fig:k478p_train}(a).
The energy differences $\omega_{74},\omega_{84}$ are a little larger than the spectral separation between harmonics. 
The interference between $4p$ and $7p/8p$ contributions probes the phase relation between next-neighbor and next-next-neighbor harmonics [see overlaps of the brown curve with the blue/red curves in Fig.~\ref{fig:k478p_train}(a)]. 
The overlap between the $7p$ and $8p$ contributions are almost perfect as their energy difference, $\omega_{7p,8p}$, is smaller than the spectral width of the harmonics, $\delta\omega$. 
This interference probes, therefore, the phase relation within each harmonic.

Figure~\ref{fig:k478p_1pulse}(b) shows the photoelectron spectrum, $P(\epsilon,\tau)$ for the attosecond pulse train, which is ionizing the 3-level Rydberg wavepacket.
The three main delay-independent signals correspond to the ionization of the $4p$ electron by the three harmonics.
The contributions of $7p$ and $8p$ and their interference cannot be spectrally distinguished. 
Their interference results in three distinct low-frequency oscillations seen in Fig.~\ref{fig:k478p_train}(b).

The $4p$--$7p/8p$ oscillations are located near 73.7~eV and 75.3~eV [see Fig.~\ref{fig:k478p_1pulse}(c)], and correspond to the interference of the 48\textsuperscript{th} with the 49\textsuperscript{th} harmonics and the 49\textsuperscript{th} with the 50\textsuperscript{th} harmonics, respectively.
A third much weaker oscillation is located at 74.5~eV. 
It corresponds to the interference between the 48\textsuperscript{th} and 50\textsuperscript{th} harmonic in agreement with Fig.~\ref{fig:k478p_train}(a). 

The extracted phases from all three oscillations [see Fig.~\ref{fig:k478p_train}(d)] agree perfectly with the analytic predictions, and Eq.~\eqref{eq:phiX-relations} hold for kinetic energies where all three beating patterns have a non-vanishing signals.
As mentioned, from the $7p$--$8p$ oscillation we can reconstruct the phase within each harmonic up to a constant.
The violet curve shows that harmonic 48 has no chirp (derivative is 0), harmonic 49 has a linear chirp, and harmonic 50 possesses a quadratic chirp.

The relative phases between harmonics can be found by analyzing the $\omega_{4p,7p}$ and/or $\omega_{4p,8p}$ beating patterns.
To determine the phase between two neighboring harmonics, the phase at only one specific kinetic energy of the $4p$--$7p$ or $4p$--$8p$ oscillations needs to be found.
At 73.7~eV and 75.3~eV, the phase differences of the $4p$--$7p$ oscillation are $-0.34\pi$ and $0.49\pi$, respectively [see red dot markings in Fig.~\ref{fig:k478p_train}(d)].
Using the knowledge about the spectral phase within each harmonics, we find that the constant phase difference between the 48\textsuperscript{th} and 49\textsuperscript{th} harmonics is $-\pi/4$, and between 49\textsuperscript{th} and 50\textsuperscript{th} harmonics is $\pi/3$ as we defined above.
Up to a constant global phase, we fully reconstructed the spectral phases of the attosecond pulse train. 
This shows that the new method is a useful tool for characterization of pulse trains (on both attosecond and femtosecond time scales simultaneously) and it can be used to test the response time of the established RABITT technique \cite{PaulScience2001}.


\subsection{Electronic structure}
\label{sec:results.estruc}

\subsubsection{Directional dipole phases}
\label{sec:results.estruc.directional}

As we discussed in Sec.~\ref{sec:theory} and seen in Sec.~\ref{sec:results.pulse}, the dipole phase drops out when studying the angle-integrated photoelectron spectrum.
In this section, we analyze the dipole phase dependence of the (directional) photoelectron spectrum of an electron ionized in the laser polarization direction.

\begin{figure}[t!]
  \includegraphics[width=\linewidth]{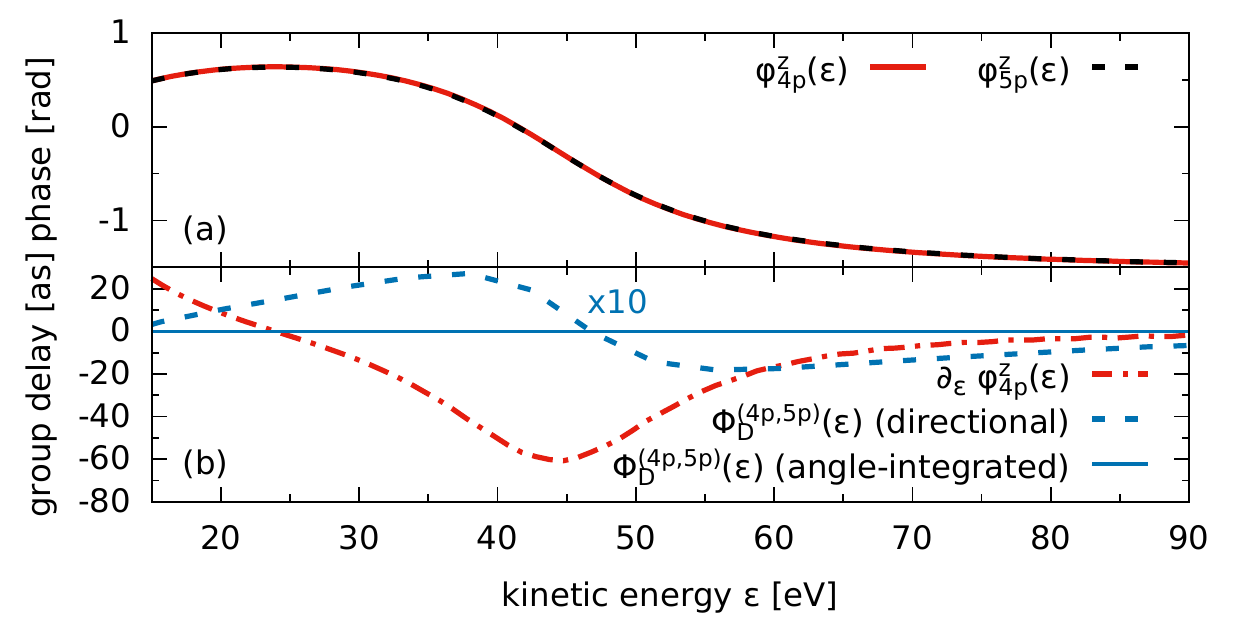}
  \caption{
  (a) Dipole phases $\varphi_{4p/5p}^z(\epsilon) = \text{arg}[\bra{\epsilon,\vec e_z} z \ket{4p/5p}]$ for ionizing an electron in $z$-direction with kinetic energy $\epsilon$.
  (b) Delay of the directional dipole phase $\varphi_{4p}^z(\epsilon)$ (red dashed line) and the relative delay between $4p$ and $5p$, $\phi_D^{(4p,5p)}(\epsilon)$, entering the directional (blue dashed line) and angle-integrated (blue solid line) photoelectron spectra. Data is produced by static HF calculations. 
  } 
  \label{fig:k45p_cm} 
\end{figure}

In Fig.~\ref{fig:k45p_cm}(a) we show the dipole phases $\varphi_{4p/5p}^z(\epsilon) = \text{arg}[\bra{\epsilon,\vec e_z} z \ket{4p/5p}]$ for ionizing an electron in $z$-direction with kinetic energy $\epsilon$.
The dipole phases vary non-monotonically by $\pi/2$ over an energy region from 20--90~eV.
This behavior can be attributed to a Cooper minimum in the photoionization cross-section of K from the $np$ states. 
It is interesting that the dipole phase of the Rydberg states $4p$ and $5p$ of potassium seem to be exactly the same.s
What enters in our scheme is, however, the marginal phase differences between the two dipoles!

The difference of the $4p$ and $5p$ dipole phases [i.e., $\phi_D^{(4p,5p)}(\epsilon)=\varphi_{4p}^z(\epsilon) - \varphi_{5p}^z(\epsilon)$ is the quantity that enters in our proposed method [see blue dashed line in Fig.~\ref{fig:k45p_cm}(b)].
The delay resulting from the phase difference is an order of magnitude smaller than the delay induced by either dipole, $\partial_\epsilon \varphi^z_{4p/5p}(\epsilon)$ (red dashed-dotted line), which enters in FROG-CRAB, PROOF and RABITT (in addition to IR-induced delays---not discussed here).
The reason for this difference is that the intermediate states in our proposed method are bound states instead of continuum states.
This shows even without angle-integration, the influence of the dipole phase is strongly reduced thanks to the similarity between neighboring Rydberg states. 
The angle-integrated phase difference, $\phi_D^{(4p,5p)}(\epsilon)$, (blue solid line) is exactly zero as discussed in Sec.~\ref{sec:theory}.

\subsubsection{Residual correlation effects}
\label{sec:results.estruc.corr}

\begin{figure}[t!]
  \includegraphics[width=\linewidth]{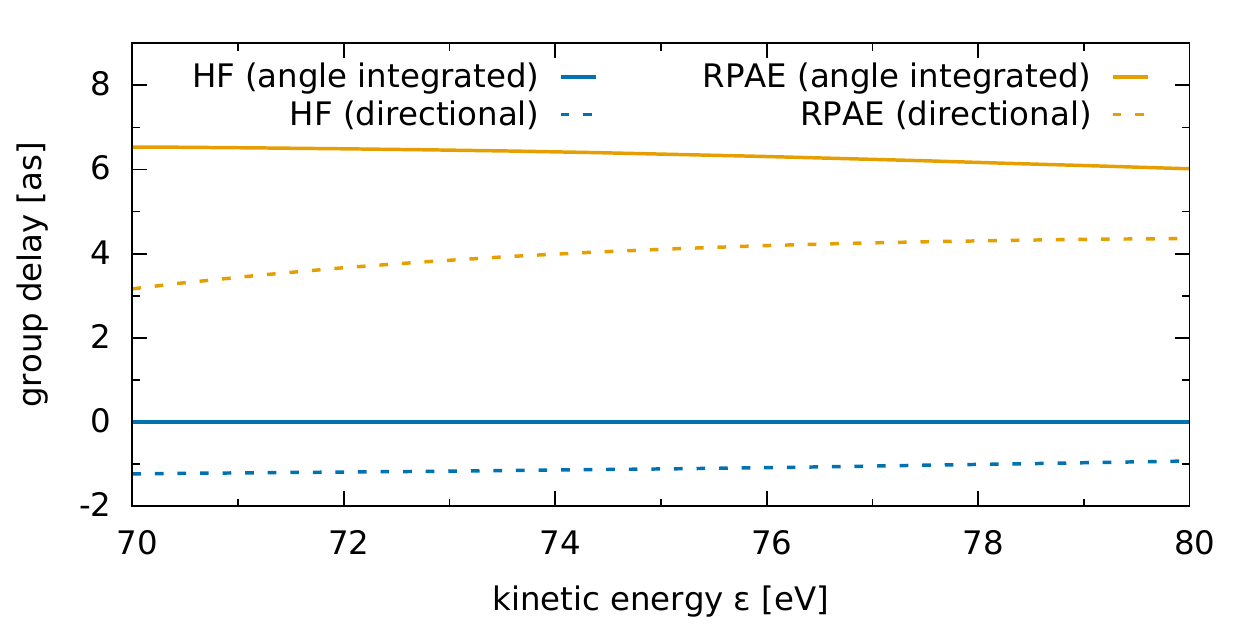}
  \caption{
  Dipole-induced delay for the directional (dashed lines) and for the angle-integrated (solid lines) photoelectron spectra of potassium treated in the HF approximation (blue lines) and RPAE (yellow lines).  
  } 
  \label{fig:k45p_direct-corr} 
\end{figure}

\begin{figure}[t!]
  \includegraphics[width=\linewidth]{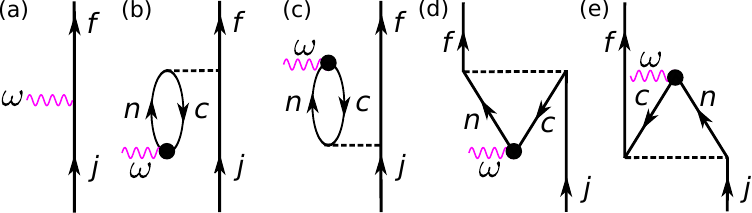}
  \caption{
Perturbation diagram for (a) HF, (b) direct-forward RPAE (c) direct-backward RPAE (d) exchange-forward RPAE  and (e) exchange-backward RPAE. Up (down) arrows label electron (hole) states. Further details are given in the main text.   
  } 
  \label{fig:arrowsRPAE} 
\end{figure}

Once electronic correlations are considered, the dipole phase affects the photoelectron spectrum even after angle integration.
Photoionization of alkali atoms is a prototypical test-case for correlation effects for close-to-threshold photoionization~\cite{ZatsarinnyPRA2010}.
In this work we consider larger photon energies where, in general, correlation effects are expected to be smaller.
In Fig.~\ref{fig:k45p_direct-corr}, we compare the retrieved phases of the $\omega_{4p,5p}$ oscillations of the directional (dashed lines) and angle-integrated (solid lines) photoelectron spectra of potassium treated within the Hartree-Fock approximation (HF; blue lines). This calculation builds on the perturbation diagrams shown in Fig.~\ref{fig:arrowsRPAE}~(a), where $j=4p, \, 5p$ are the excited Rydberg states and $f=ks, \, kd$ are the final photoelectron states, both computed within HF. 
This type of calculation contains no correlation effects by definition and the group delay of the angle-integrated calculation is zero, as expected from Sec.~\ref{sec:theory}. 
In Fig.~\ref{fig:k45p_direct-corr} we also show the random-phase approximation with exchange (RPAE; yellow lines), which includes additionally correlation effects with the electrons from the $Ar^+$ core $c=3s, \, 3p$, as illustrated by the perturbation diagrams in Fig.~\ref{fig:arrowsRPAE}~(b)--(e). In Fig.~\ref{fig:arrowsRPAE}~(b) an UV photon (wiggly line) creates a hole $c$ in the core and a virtual electron $n$ that then annihiliate by Coulomb interaction (dashes line) by kicking out the Rydberg electron $j \rightarrow f$. The RPAE calculation includes exchange Fig.~\ref{fig:arrowsRPAE}~(d) and backward diagrams (c) and (e). It is correct to first order in Coulomb interactions, but it includes an infinite summation of partial diagrams indicated by a solid dot in Fig.~\ref{fig:arrowsRPAE}. More details about RPAE are found in Ref.~\cite{Amusia1990}.  

The direction of the photoelectron is chosen to be parallel to the laser polarization direction ($\theta=0$).
The energy range of 70--80~eV is chosen to avoid on-shell excitation of autoionizing resonances in the continuum (see Sec.~\ref{sec:results.estruc.fano}) and to focus on the residual effect (virtual coupling) to other particle-hole excitations.

When correlation effects are included, the reduced dipole moments become complex-valued even in spherically symmetric systems resulting in a dipole phase dependence in the angle-integrated photoelectron spectrum.
In Fig.~\ref{fig:k45p_direct-corr}, the RPAE results, which include correlation, show a weak energy dependence.
The correlation induced delay in the angle-integrated spectrum (solid yellow line) is centered around 6~as.
A constant delay difference is, however, not of interest and does not influence the determination of the spectral chirp.
The energy dependence is thanks to the Rydberg states again quite weak.
The average slope is -0.052~as/eV in the 70--80~eV range.
The influence of correlation on the directional photoelectron (dashed yellow line) is with an average slope of 0.12~as/eV more than a factor of two stronger than in the angle-integrated result and has a different sign in the slope.
It shows even in the presence of correlations, the dipole phase dependency is significantly reduced, when studying the angle-integrated photoelectron spectrum, and the accuracy of the reconstructed spectral chirp is greatly enhanced.

The dipole dependence due to correlation can be further reduced when going to higher Rydberg states, or when going to lighter atoms (e.g. sodium with neon-like core), which generally contain less electronic correlation.

\subsubsection{Fano resonances}
\label{sec:results.estruc.fano}

When the ionizing test pulse is resonant with autoionizing inner-shell excitations, Fano resonances become visible in the photoelectron spectrum, which are spectrally highly localized around the autoionizing resonance energy (in contrast to the residual effects discussed in Sec.~\ref{sec:results.estruc.corr}).
Since autoionization is a correlated process, the dipole phase dependence will survive in the photoelectron spectrum after angle integration in the form of amplitude and phase modulations in the beating patterns.

To demonstrate the influence of an autoionizing state, we choose neon and we target the lower $2s^{-1}ns$ autoionizing states, which are even-parity states and can only be reached with an even number of photons.
First, we prepare a Rydberg wavepacket between $2p^{-1}3s$ and $2p^{-1}4s$ with two 4.8~fs ($=200$~a.u.) Gaussian pulses with center frequencies 16.8~eV and 19.7~eV.
After the wavepacket is prepared, we ionize it with a 508~as ($=21$~a.u.) Fourier-limited Gaussian pulse with a center frequency of 25.9~eV ($=0.95~E_h$), which ionizes the Rydberg electron but also drives the ionic transition between $2p^{-1}$ and $2s^{-1}$ leading to population of the autoionizing $2s^{-1}ns$ states. 
The calculations~\footnote{
A pseudo-spectral grid with a radial box size of 220~$a_0$, 1000 grid points, and a mapping parameter of $\zeta=0.5$ are used. 
The splitting function starts around $R_\text{split}=170~a_0$, a smoothness of 2~$a_0$, and splitting interval of $dt_\text{split}=1$~a.u. 
The maximum angular momentum is 3 and Hartree-Fock orbitals up to an energy of 10~$E_h$ are considered. 
The propagation method is Runge-Kutta 4 with a time step $dt=0.05$~a.u.
}
are based on time-dependent configuration-interaction singles (TDCIS)~\cite{GrSa-PRA-2010}, which accurately describes the $2s^{-1}np$ Fano resonances in neon~\cite{HePa-PRA-2014}.

\begin{figure}[t!]
  \includegraphics[width=\linewidth]{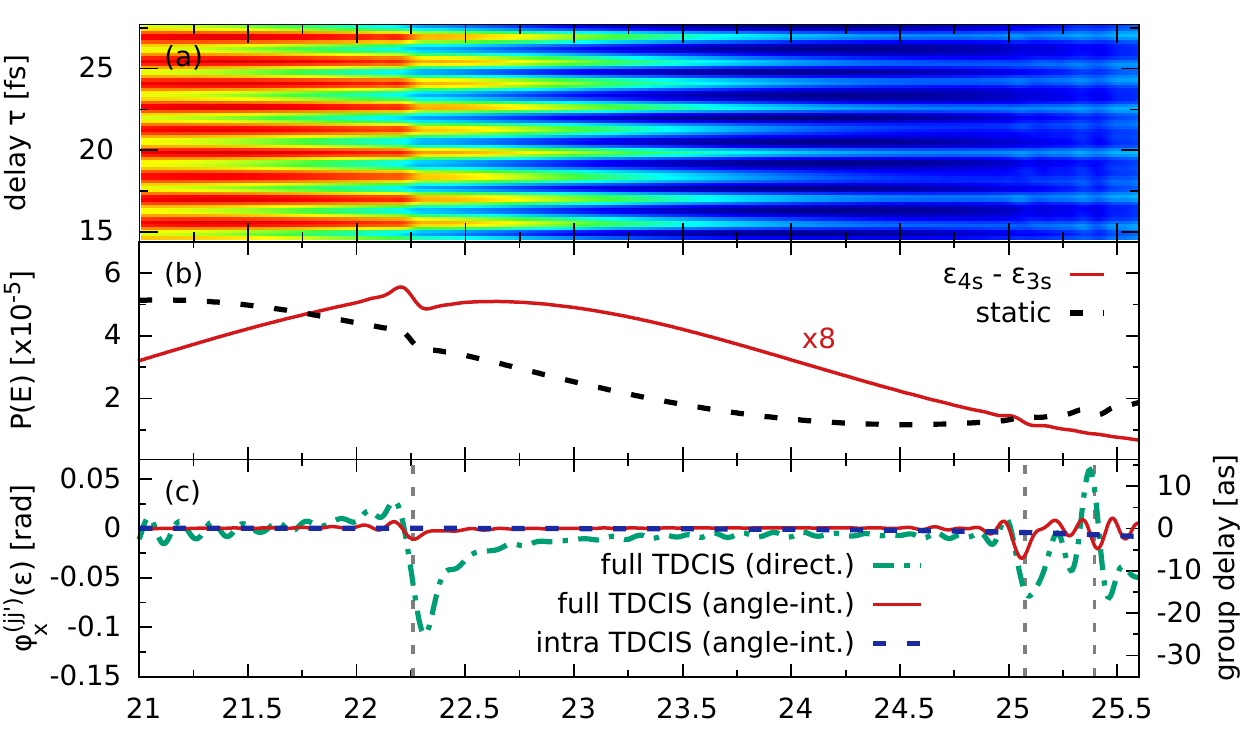}
  \caption{
  (a) Angle-integrated photoelectron spectrum of the $2p^{-1}3s$--$2s^{-1}4s$ wavepacket ionized by an isolated attosecond pulse.
  (b) The static (black dashed line) and the oscillating components (solid red line) of the angle-integrated photoelectron spectrum.
  (c) Retrieved phase of the oscillating components of the angle-integrated (red solid line) and directional (green dashed-dotted line) photoelectron spectrum for the full TDCIS model. 
  The intrachannel TDCIS results for the angle-integrated photoelectron spectrum is shown as reference (blue dashed line).
  The corresponding group delay of the phase differences is shown on the right.
  }
  \label{fig:ne_fano} 
\end{figure}

Phase modulations in the beating pattern will be only due to $\phi_D^{jj'}(\epsilon_f)$ because $\Phi_X^{jj'}(\epsilon_f)=0$ in Eq.~\ref{eq:Theta-directional}.
In Fig.~\ref{fig:ne_fano}(a), the angle-integrated photoelectron spectrum around the $2s^{-1}3s$ is shown and Fig.~\ref{fig:ne_fano}(b) shows the amplitude of the static and oscillating contributions. 
The $2s^{-1}3s$ and higher Fano-resonances are visible in the static and especially the oscillating amplitudes.
The static signal increases beyond 25~eV due to two-photon absorption of the attosecond pulse. 
This process is not delay-dependent and contributes only to the static background.

The retrieved phases are shown in Fig.~\ref{fig:ne_fano}(c) where the energy position of the $2s^{-1}ns$ resonances are highlighted by vertical dashed lines. 
Also the intrachannel TDCIS result for the angle-integrated photoelectron spectrum is shown (blue dashed line), where the interchannel interactions are responsible for the autoionization of all $2s^{-1}nl$ states are not included. 
Ignoring interchannel effects eliminates correlation effects, and as a result no phase modulations around the resonance energies are seen. 
The phase changes due to the $2s^{-1}3s$ Fano resonance is quite small for the angle-integrated photoelectron spectrum.
The induced delay does not exceed $\pm 3$~as. 
For the directional photoelectron spectrum, the correlation induced delay is with up to -30~as an order of magnitude larger than for the angle-integrated result.
This shows as in Sec.~\ref{sec:results.estruc.corr}, angle-integration reduces the influence of the dipole phase dramatically even in the presence of correlations. 
However, in both cases (directional or angle-integrated), the effect in the delay is relatively small compared to the $2p^{-1}3s$ resonance lifetime of 6.4~fs ($\Gamma = 0.1$~eV within TDCIS).
The derivative of the absolute dipole phase, which would be measured in FROG-CRAB would reflect the lifetime of the resonance and result in large phase corrections.
In our method, we measure the difference in the influence of the $2s^{-1}3s$ autoionizing state on the Rydberg states $2p^{-1}3s$ and $2p^{-1}4s$ resulting in a correlation-induced delay, which is at least two orders of magnitude smaller.

\section{Conclusion}
\label{sec:conclusion}

We have proposed a novel method to characterize attosecond UV pulses with the help of bound electron wavepackets. Different spectral components of the UV pulse interfere as an electron can be ionized from different energy levels. This leads to quantum beats (an oscillating photoelectron signal) as function of the pulse delay. We showed that angle integration of the photoelectron spectrum eliminates the influence of the final state scattering phase.  

Rydberg wavepackets have favorable properties for the proposed method, in particular, the energy spacing between electronic states decreases with $1/n^3$ offering high spectral energy resolution. 
We showed that a wavepacket consisting of multiple levels can be used for characterization pulse trains. 
Multi-level wave packets also opens up for consistency checks of the retrieved phases---a feature that does not exist in any established pulse reconstruction technique.

We studied the role of correlation effects and we found that, while these effects are small, they cannot be completely eliminated by angle-integration of the photoelectron. Using Rydberg wavepackets minimized this effects.
We have shown in potassium and neon correlation effects result indeed in negligible phase corrections.
The possibility to use vibrational wavepackets with a subsequent photodissociation step offers exiting opportunities to study correlated non-Born-Oppenheimer dynamics. 

We believe ionizing Rydberg wavepackets is a versatile approach to determine, with unprecedented precision, the spectral phases of complex pulses, e.g. generated by table-top attosecond laboratories or FEL facilities. We have focused on pulses in the UV. This technique can be easily extended to x-rays with shorter wavelengths and into the optical and near-infrared domains with longer wavelengths.
As pulses durations approach the zeptosecond regime~\cite{PoCh-Science-2012}, ionization from different inner shells cannot be distinguished anymore making FROG-CRAB and PROOF more prone to error.
Our method is not affected by inner-shell ionization, and the high spectral phase accuracy and the applicability to a very broad spectral range makes our method ideal for characterizing ever shorter pulses. 

%
%

\begin{acknowledgments} 
We thank Eva Lindroth for stimulating discussions.
We acknowledge the support of the Kavli Institute of Theoretical Physics (National Science Foundation under Grant No. NSF PHY11-25915)
and NORDITA for support during the workshop on 'Control of Ultrafast Quantum Phenomena'. 
S.P. is funded by the Alexander von Humboldt Foundation and by the NSF through a grant to ITAMP.
J.M.D. is funded by the Swedish Research Council, Grant No. 2014-3724. 
\end{acknowledgments}

%

\bibliography{references,amo,books}

\end{document}